\documentclass[12pt,a4paper]{article}
\usepackage{amsmath}
\usepackage{graphicx}
\usepackage{times}
\begin{document}
\begin{titlepage}
\title{On the geometric phase and the scattering at the LHC}
\author{ S.M. Troshin, N.E. Tyurin\\[1ex]
\small  \it Institute for High Energy Physics,\\
\small  \it Protvino, Moscow Region, 142281, Russia}
\normalsize
\date{}
\maketitle

\begin{abstract}
We discuss appearance at the LHC energies of the geometric phase, which in its turn can reflect a presence  of the non-factorizable background geometry
proposed in the RS-scenario with extra spatial dimensions.
\end{abstract}
\end{titlepage}
\setcounter{page}{2}

An issue of the geometric phase in the high energy scattering is discussed.  In quantum mechanics  the geometric or Berry phase  appears
 as a result of a time evolution resulting in the adiabatic loop (cyclic) variation of the Hamiltonian parameters \cite{praman,berry}. 
 This phase was originally discovered 
for the states corresponding to the discrete spectrum and for the case of continuous spectrum in 
quantum mechanics the issue of the geometric phase has been studied in \cite{newton}. This adibatic loop evolution of the Hamiltonian parameters
could lead to the phase difference of the initial and final states which is independent of the system dynamics and, in fact, depends only on the path 
the system evolves along of.
Such phase difference is known as the Berry phase
 and its appearance  demonstrates  singular parameter dependence of the system dynamics. 
The interesting phenomenon  of the geometric phase appearance can be found in many physical systems \cite{bphase}.

In the relativistic high energy scattering Hamiltonian is not known; instead, one should deal with the scattering matrix. 
Here, we consider the issue of the geometric phase in the case of relativistic high energy scattering on the base of the unitary representations for
the scattering matrix and provide  new details in addition to the ones  mentioned in \cite{reflect}.

 The elastic scattering $S$-matrix is connected with the elastic scattering amplitude by the relation $S(s,b)=1+2if(s,b)$, which is written here
in the impact parameter representation. At high
energies the respective function $S(s,b)$ can be rewritten in the form
\begin{equation}\label{ssb}
 S(s,b)=\kappa(s,b)\exp[2i\delta(s,b)],
\end{equation}
where $\kappa(s,b)$ and $\delta(s,b)$ are the real functions, $b$ is an impact parameter of colliding hadrons. The function $\kappa$  
($0\leq \kappa \leq 1$)
 is an absorption factor
related to the contribution of the inelastic channels into unitarity relation, its zero value, $\kappa=0$, corresponds to a complete 
absorption of the incoming channel. 
 The factor  $\kappa(s,b)$ is determined by the inelastic channels contribution
to the unitarity equation for the elastic scattering amplitude $f(s,b)$, i.e.
\begin{equation}\label{kap}
 \kappa^2(s,b)=1-4h_{inel}(s,b),
\end{equation}
with $h_{inel}(s,b)$ entering the  equation
\begin{equation}\label{unit}
 \mbox{Im}f(s,b)=h_{el}(s,b)+h_{inel}(s,b).
\end{equation}

When elastic scattering amplitude is a pure imaginary function, the function $S(s,b)$ is a real one
and it can vary in the range $-1<S(s,b)<1$. Evidently,  at fixed value of $s$ the function
 $S(s,b)\to 1$ at $b\to\infty $, i.e. in this limit $\kappa \to 1$ and $\delta \to 0$.

Let 
consider another limiting behavior of $S(s,b)$, namely, when the impact parameter $b$ is fixed and $s\to \infty$. 
The standard assumption is that the function $S(s,b)\to 0$ in this limit. It corresponds to
approaching a black disk limit  and the elastic scattering  then is just a shadow of all
inelastic processes. However, there is no reason to exclude existence of an another option, i.e. the function $S(s,b)\to -1$ when $b$ is fixed 
and $s\to \infty$, i.e.  $\kappa \to 1$ and $\delta \to \pi/2$. This limiting behavior  corresponds to a reflective scattering \cite{reflect}. 
It implies appearance of the non-zero phase $\delta$ equal to $\pi/2$. 

In the relativistic scattering one can consider the absorption factor $\kappa$  as a parameter which determine the hadron interaction dynamics at high energies. 
We can force  $\kappa$ to perform a loop variation similar to a loop variation of the Hamiltonian parameters in quantum mechanics.  
Namely, we can  vary variable $s$ (and/or $b$) in the way
 that $\kappa$ 
  evolves cyclically and adiabatically from $\kappa_i>0$ to $\kappa_{min}=0$ and
  then to the value $\kappa_f$, where $\kappa_f=\kappa_i$ (i.e. performing a loop variation).
  As a result
 the  non-zero phase appears ($\delta=\pi/2$) at small values of the impact parameter and this phase
  is independent of the details
   of the energy and/or impact parameter evolution. Therefore, we can  treat it as a geometric phase, considering it as an analog of the Berry phase in quantum mechanics.

To illustrate the evolution which can lead to this loop variation one can use rational form of $S$--matrix unitarization 
which provide antishadowing in the elastic scattering \cite{phl}. This approach can be considered as a resulting one from the
confinement condition implication for the colored states \cite{coll}.
In the $U$--matrix approach 
 the elastic scattering matrix element in the
impact parameter representation
is the following linear fractional transform:
\begin{equation}
S(s,b)=\frac{1-U(s,b)}{1+U(s,b)}. \label{um}
\end{equation}
 $U(s,b)$ is the generalized reaction matrix, which is considered to be an
input dynamical quantity. This form (\ref{um}) is one-to-one transform and is easily
invertible. We consider pure imaginary scattering amplitude $f$ and it leads to the pure imaginary
function $U$.
The absorption factor $\kappa(s,b)$
is connected to the function $U(s,b)$ by the relation
\begin{equation}
\kappa(s,b)=\left|\frac{1- U(s,b)}{1+U(s,b)}\right|\label{uf}.
\end{equation}
The  most of the models \cite{reflect,gidd} provide increasing
dependence of the function $U(s,b)$ with energy (e.g. power-like one) and the exponential decrease
with impact parameter at large values of $b$.  The most recent quantitative  analysis of the available experimental data with this form of unitarization 
has been performed in \cite{mart}.
The value of energy corresponding to the black disk realization
in central collisions  $S(s,b)|_{b=0}=0$
will be denoted as $s_0$ and it is determined by the  equation
$U(s,b)|_{b=0}=1$; there is a model estimate for $\sqrt{s_0}$ at around 2 $TeV$ \cite{phl}.
In the energy region $s\leq s_0$ the scattering in the whole range of impact parameter variation
has a pure shadow nature and the elastic $S$-matrix varies in the range
$0\leq S(s,b)<1$. But when the energy is higher than $s_0$
the scattering picture at small values of impact parameter $b\leq R(s)$
aquires  geometric (of reflective) nature while shadow contribution is decreasing with energy,  $R(s)$ (nodal point)
is determined by $U(s,b=R(s))=1$. 
The $S$-matrix varies in the region  $-1<S(s,b)\leq 0$ at $s\geq s_0$
at $b\leq R(s)$. 
The  impact parameter dependence of the absorption factor $\kappa(s,b)$ and phase $\delta(s,b)$  is depicted on Fig. 1
 for the case of $s>s_0$.
\begin{figure}[h]
\begin{center}
\resizebox{12cm}{!}{\includegraphics*{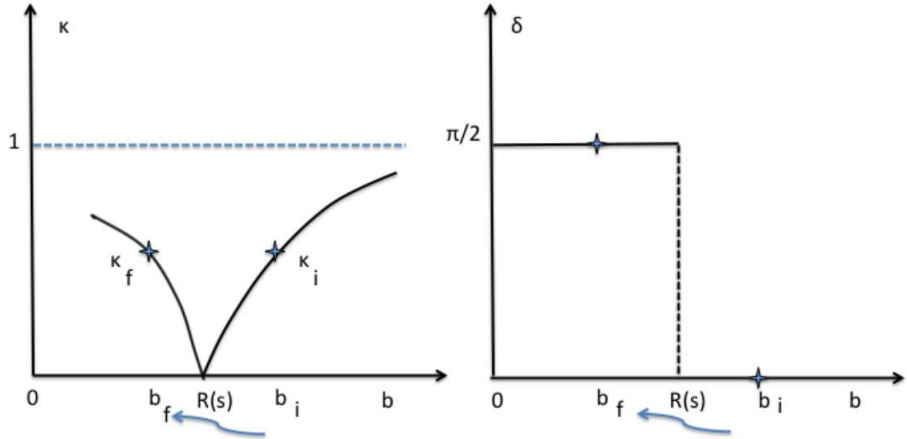}}
\end{center}
\caption{Appearance of the geometric phase at the energies $s>s_0$ under the change of the impact parameter from $b_i$ to $b_f$ with $\kappa_i=\kappa_f$, 
where $\kappa_{i,f}\equiv\kappa(s,b_{i,f})$.}
\end{figure}
Thus, we can observe that despite the factor $\kappa$ performs loop variation when initial and final values are the same ($\kappa_i=\kappa_f$),
the non-zero geometric or Berry phase appears. 

One might relate it with the manifestation of the extra dimension with non-factorized metrics proposed
in the RS-scenario \cite{rs}.
There are two different realization of the idea of the extra dimensions: one (ADD-model \cite{add})  with a metrics of the
 compact extra dimensions factorized out of the metrics of the 
standard 3+1 manifold. In another theory (RS-model) called 5-dimensional warped geometry model \cite{rs}  the space
 is a five-dimensional anti-de Sitter space with non-factorized metrics. 
The issues of extra dimensions are very well developed now and there are a lot of publications devoted to this subject (cf.  review paper \cite{ferug} 
and the references therein).  Here we just comment on the 
the difference between the scenarios with factorizable and non-factorizable geometries. Namely, one should expect that the geometric phase
will be presented in the latter case and  absent in the former one.

We discuss now a possible instrumentation  for the selection
of the observable effects related to the presence of the geometric phase. 
Namely, for that purpose,  it is useful to consider the ratio of elastic to total  impact parameter dependent
 cross-sections, i.e.
\[
 {\cal R}(s,b)=\sigma_{el}(s,b)/\sigma_{tot}(s,b).
\]
The impact parameter dependent cross--sections $\sigma_{el}(s,b)$, $\sigma_{inel}(s,b)$ and $\sigma_{tot}(s,b)$ can be extracted from the
experimental data as it was performed in \cite{ama}.
The function ${\cal R}(s,b)$ at the energies $s>s_0$ has two different forms depending on the value of impact parameter, namely at $b>R(s)$ this function has the form
\[{\cal R}(s,b)=[1-\kappa(s,b)]/2,\] while in the region $b<R(s)$ the presence of the geometric phase (i.e. $\cos 2\delta=-1$)  changes this 
dependence to \[{\cal R}(s,b)=[1+\kappa(s,b)]/2\] 
and it means that the hadron scattering is becoming predominantly  geometric one at small impact parameters and high energies $s>s_0$. 
Thus, the scattering has a significant absorptive contribution only 
in the peripheral region of the impact parameters. The  following asymptotic limits take place, namely,   ${\cal R}(s,b)\to 1$ when 
 impact parameter $b$ is fixed and $s\to\infty$ 
(i.e. geometric elastic scattering
saturates the total cross-section in this limit), 
while ${\cal R} (s,b)\to 0$ at fixed energy $s$
 and $b\to\infty$, i.e.  at large impact parameters elastic scattering cross-section decreases faster than the cross-sections of the all
inelastic processes, i.e. 
\[
 {\cal R}(s,b)={\bar{\cal R}}(s,b)/[1+{\bar{\cal R}}(s,b)],
\]
where
\[{\bar{\cal R}}(s,b)\equiv \sigma_{el}(s,b)/\sigma_{inel}(s,b)\to 0\]
 in this limit. This is evident since
\[{\bar{\cal R}}(s,b)={f(s,b)}/{[1-f(s,b)]}\] and $f(s,b)\to 0$ at fixed $s$ when $b\to\infty$. It is also clear
that \[{\bar {\cal R}}(s,b)\to \infty\] at fixed $b$ and $s\to\infty$ in the presence of the above geometric phase.

It should be noted that the ratio ${\bar{\cal R}}(s,b)$ determines the function $U(s,b)$, i.e.
 \[ U(s,b)={\bar{\cal R}}(s,b) \]
and can be used as a tool to obtain a qualitative behavior of the function $U(s,b)$
from the experimental data\footnote{It should be  reminded here that we consider for simplicity a  case of  pure imaginary scattering amplitude. It  helps to separate
the effects related to the geometric phase from the ones associated with the dynamical phase.}.

The aim of this  note was to discuss an issue of the geometric phase in the high energy scattering and to point out to the possible relation of the geometric phase 
with the presence of the extra dimensions with non-factorizable metrics along with discussion of 
the relation of this phase to some experimental facts observed or predicted at the LHC energies. 

The  appearance of the geometric phase at $s>s_0$ provides a change in the  slope (''the knee'') of the energy spectrum of the cosmic particles 
observed at the ground level
as a result of  the cosmic particles interactions with the air  \cite{reflect}. 

It is the presence of geometric phase at the LHC energies means that geometric elastic scattering becomes dominating in the region of small impact parameters while inelastic
processes are predominantly peripheral.  It means  that albedo is increasing with energy 
in the high energy hadron scattering at $s>s_0$  \cite{reflect}.

The perepherality of the inelastic amplitudes would lead to the ridge and double-ridge effects the the two-particle correlation functions 
in proton-proton collisions \cite{coll,dri}. 

We also note that the presence of this phase should be interpreted in favor 
of the RS-scenario of the extra dimensions modeling.

\end{document}